# LMI-based robust stability and stabilization analysis of fractional-order interval systems with time-varying delay

Pouya Badri[1], Mahdi Sojoodi[1*]

[1]Advanced Control Systems Laboratory, School of Electrical and Computer Engineering, Tarbiat Modares University, Tehran, Iran.
[*]sojoodi@modares.ac.ir

**Abstract:**
This paper investigates the robust stability and stabilization analysis of interval fractional-order systems with time-varying delay. The stability problem of such systems is solved first, and then using the proposed results a stabilization theorem is also included, where sufficient conditions are obtained for designing a stabilizing controller with a predetermined order, which can be chosen to be as low as possible. Utilizing efficient lemmas, the stability and stabilization theorems are proposed in the form of LMIs, which is more suitable to check due to various existing efficient convex optimization parsers and solvers. Finally, two numerical examples have shown the effectiveness of our results.

## 1. Introduction

In the last decades, utilizing fractional-order calculus has open new horizons in modeling real-world systems, since it can more concisely describe the behavior of systems having a response with long memory transients and other intrinsic features which are more suitable with fractional-order equations [1-4]. Moreover, it has been proven that fractional-order controllers have significant advantages over integer-order ones [2, 5]. Hence, a lot of studies have been focused on the fractional-order control systems [3, 6-9].

Real systems in various areas, such as engineering, biology, and economics are sometimes confronted with time delays, which can lead to instability and oscillations in such systems [1, 6, 7]. Thus, in the past few years, stability and stabilization problem of time-delay systems has attracted particular attention [10], including fractional-order systems [1, 11-14]. In [15] the robust stabilization problem of interval fractional-order systems with one time-delay using fractional-order controllers by the Minkowski sum of value sets was investigated. Moreover, in [1] the robust stability of fractional-order interval systems with multiple time delays was discussed. In [13], the robust stability of a fractional-order time-delay system is investigated in the frequency domain based on finite spectrum assignment. This algorithm is an extension of the traditional pole assignment method, which can change the undesirable system characteristic equation into a desirable one. Furthermore, in [14], a robust fractional-order PID controller is designed for fractional-order delay systems based on positive stability region (PSR) analysis. The finite-time stability of linear delay fractional-order systems is also investigated in [16] based on the generalized Gronwall inequality and the Caputo fractional derivative.

Moreover, modeling real-world processes usually leads to uncertain models due to neglected dynamics, uncertain physical parameters, parametric variations in time, and so on. Therefore, robust stability and stabilization became an important problem for all control systems including fractional-order ones [15, 17-19]. In [18] the problems of the robust stability and stabilization of fractional-order linear systems with positive real uncertainty are proposed. As it has been declared that interval uncertainty is more convenient for the control system design problems [20] and robust stability analysis [21], stability and stabilization problems of fractional-order interval system are investigated in [17]. A new sufficient condition in terms of LMI for the global asymptotic stability of a class of interval fractional-order nonlinear systems with time-varying delay was proposed in [12], where the state matrix of the linear part of the system is supposed to be diagonal.

In the majority of available controller design methods, high-order controllers are obtained suffering from costly implementation, high fragility, unfavorable reliability, maintenance difficulties, and potential numerical errors. Designing a controller with a low and fixed-order would be helpful because the desired closed-loop performance is not necessarily assured by available plant or controller order reduction procedures [22]. Therefore, in our previous works, fixed-order controllers have been designed for fractional-order systems [3, 4, 19].

Motivated by aforementioned observations, our paper aims at solving the problem of stability and stabilization of interval fractional-order systems with time-varying delay in terms of linear matrix inequalities LMIs, which is suitable to be used in practice due to various efficient convex optimization parsers and solvers that can be applied to determine the feasibility of the LMI constraints and consequently calculate design parameters.

The main contributions of this paper can be summarized as follows:
- LMI conditions are obtained for stability check of fractional-order interval systems with time-varying delay.
- Robust stabilizing problem of such systems is investigated using proposed LMI stability conditions.
- Fixed-order dynamic output feedback controller order is designed whose order can be determined before design.

As far as we know, there is no result on the robust stability of uncertain FO-LTI systems, with time-varying delays in the literature. Moreover, analytical design of a stabilizing dynamic output feedback controller for interval fractional-order systems with time-varying delay is investigated for the first time. It is worth noting that, LMI stability conditions for uncertain FO systems with time delay, which are more comfortable to check, are obtained for the first time in this paper.

The rest of this paper is organized as follows: In section 2, some preliminaries about interval uncertainty and fractional-order calculus together with the problem formulation are presented. LMI-based robust stability and stabilizing conditions using a dynamic output feedback controller are derived in Section 3. Some numerical examples are given in Section 4 to illustrate the effectiveness of the proposed theoretical results. Finally, the conclusion is drawn in section 5.

**Notations**: In this paper, by $M^T$ we denote the transpose of matrix $M$, and $Sym(M)$ denotes $M + M^*$. The notation $\bullet$ is the symmetric component symbol in matrix and $\uparrow$ is the symbol of pseudo inverse. Moreover, The notations **0** denotes the zero matrix with appropriate dimensions.

## 2. Preliminaries and problem formulation

In this section, some basic concepts and lemmas of fractional-order calculus and interval uncertainty are presented.

Consider the following uncertain FO-LTI system for $0 < \alpha < 1$:

$$\begin{cases} D^\alpha x(t) = Ax(t) + Bu(t - d(t)), t > 0, \\ x(t) = \phi(t), \quad t \in [-\tau, 0] \end{cases} \tag{1}$$

in which $x(t) \in R^n$ denotes the pseudo-state vector, $u \in R^l$ is the control input, and $y \in R^m$ is the output vector. Furthermore, $A \in R^{n \times n}$ and $B \in R^{n \times l}$ are interval uncertain matrices as follows

$$A \in A_I = [\underline{A}, \overline{A}] = \{[a_{ij}]: \underline{a}_{ij} \le a_{ij} \le \overline{a}_{ij}, 1 \le i, j \le n\}, \tag{2}$$

$$B \in B_I = [\underline{B}, \overline{B}] = \{[b_{ij}]: \underline{b}_{ij} \le b_{ij} \le \overline{b}_{ij}, 1 \le i \le n, 1 \le j \le l\}, \tag{3}$$

where $\underline{A} = [\underline{a}_{ij}]_{n \times n}$ and $\overline{A} = [\overline{a}_{ij}]_{n \times n}$ satisfy $\underline{a}_{ij} \le \overline{a}_{ij}$ for all $1 \le i, j \le n$, $\underline{B} = [\underline{b}_{ij}]_{n \times l}$ and $\underline{B} = [\underline{b}_{ij}]_{n \times l}$ satisfy $\underline{b}_{ij} \le \overline{b}_{ij}$ for all $1 \le i \le n, 1 \le j \le l$. The time delay $d(t)$ is a time-varying continuous function that satisfies

$$0 \le d(t) \le \tau, \tag{4}$$

and

$$\dot{d}(t) \leq \mu < 1 \tag{5}$$

where $\tau$ and $\mu$ are constants and the initial condition $\phi(t)$ represents a continuous vector-valued initial function of $t \in [-\tau, 0]$.

In this article, the following Caputo definition for fractional derivatives of order $\alpha$ of function $f(t)$ is utilized [23]:

$$^C_aD^\alpha_t f(t) = \frac{1}{\Gamma(m-\alpha)} \int_a^t (t-\tau)^{m-\alpha-1} \left(\frac{d}{d\tau}\right)^m f(\tau)d\tau$$

where $\Gamma(\cdot)$ is Gamma function defined by $\Gamma(\epsilon) = \int_0^\infty e^{-t} t^{\epsilon-1} dt$ and $m$ is the smallest integer that is equal to or greater than $\alpha$.

The following notations are needed for dealing with interval uncertainties.

$$A_0 = 1/2(\underline{A} + \overline{A}), \Delta A = 1/2(\overline{A} - \underline{A}) = \{\gamma_{ij}\}_{n \times n}, \tag{6}$$

$$B_0 = 1/2(\underline{B} + \overline{B}), \Delta B = 1/2(\overline{B} - \underline{B}) = \{\beta_{ij}\}_{n \times l}, \tag{7}$$

It is evident that all elements of $\Delta A$ and $\Delta B$ are nonnegative, therefore the following matrices are defined.

$$M_A = [\sqrt{\gamma_{11}}e_1^n \quad \cdots \quad \sqrt{\gamma_{1n}}e_1^n \quad \cdots \quad \sqrt{\gamma_{n1}}e_n^n \quad \cdots \quad \sqrt{\gamma_{nn}}e_n^n]_{n \times n^2}, \tag{8}$$

$$R_A = [\sqrt{\gamma_{11}}e_1^n \quad \cdots \quad \sqrt{\gamma_{1n}}e_n^n \cdots \quad \sqrt{\gamma_{n1}}e_1^n \quad \cdots \quad \sqrt{\gamma_{nn}}e_n^n]^T_{n^2 \times n}, \tag{9}$$

$$M_B = [\sqrt{\beta_{11}}e_1^n \quad \cdots \quad \sqrt{\beta_{1l}}e_1^n \quad \cdots \sqrt{\beta_{n1}}e_n^n \quad \cdots \quad \sqrt{\beta_{nl}}e_n^n]_{n \times nl}, \tag{10}$$

$$R_B = [\sqrt{\beta_{11}}e_1^l \quad \cdots \quad \sqrt{\beta_{1l}}e_l^l \quad \cdots \sqrt{\beta_{n1}}e_1^l \quad \cdots \quad \sqrt{\beta_{nl}}e_l^l]^T_{nl \times l}, \tag{11}$$

where $e_k^n \in R^n$, $e_k^l \in R^l$, and $e_k^m \in R^m$ are column vectors with the $k$-th element being 1 and all the others being 0. In addition, we have

$$H_A = \{diag(\delta_{11}, \ldots, \delta_{1n}, \ldots, \delta_{n1}, \ldots, \delta_{nn}) \in R^{n^2 \times n^2}, |\delta_{ij}| \leq 1, i,j\ 1, \ldots, n\}, \tag{12}$$

$$H_B = \{diag(\eta_{11}, \ldots, \eta_{1l}, \ldots, \eta_{n1}, \ldots, \eta_{nl}) \in R^{(nl) \times (nl)}, |\eta_{ij}| \leq 1, i = 1, \ldots, n, j = 1, \ldots, l\}, \tag{13}$$

The following lemmas are required, to study the stability of interval fractional-order systems.

**Lemma 1** [17]: Let

$$A_J = \{A = A_0 + M_A F_A R_A | F_A \in H_A\}, B_J = \{B = B_0 + M_B F_B R_B | F_B \in H_B\}, \tag{14}$$

then $A_I = A_J$, and $B_I = B_J$.

**Lemma 2** [17]: For any matrices $X$ and $Y$ with appropriate dimensions, we have

$$X^T Y + Y^T X \leq \eta X^T X + (1/\eta) Y^T Y \text{ for any } \eta > 0. \tag{15}$$

**Lemma 3** [6]: For given scalars $\tau > 0$ and $\mu < 1$, the following certain integer-order system

$$\begin{cases} \dot{x}(t) = Ax(t) + Bx(t - d(t)), & t > 0 \\ x(t) = \phi(t), & t \in [-\tau, 0] \end{cases} \tag{16}$$

with fixed matrices $A$ and $B$ and a time-varying state delay $d(t)$ satisfying (4) and (5) is asymptotically stable if there exist $P = P^T > 0$, $Q = Q^T \geq 0$, $Z = Z^T > 0$ and appropriately dimensioned matrices $N_i$ and $T_i$ ($i = 1,2,3$) such that the following *LMI* holds:

$$\Gamma = \begin{bmatrix} \Gamma_{11} & \Gamma_{12} & \Gamma_{13} & \tau N_1 \\ \Gamma_{12}^T & \Gamma_{22} & \Gamma_{23} & \tau N_2 \\ \Gamma_{13}^T & \Gamma_{23}^T & \Gamma_{33} & \tau N_3 \\ \tau N_1^T & \tau N_2^T & \tau N_3^T & -\tau Z \end{bmatrix} < 0. \tag{17}$$

where

$$\Gamma_{11} = Q + N_1 + N_1^T - A^T T_1^T - T_1 A, \quad \Gamma_{12} = N_2^T - N_1 - A^T T_2^T - T_1 B, \quad \Gamma_{13} = P + N_3^T + T_1 - A^T T_3^T, \tag{18}$$
$$\Gamma_{22} = -(1-\mu)Q - N_2 - N_2^T - T_2 B - B^T T_2^T, \quad \Gamma_{23} = -N_3^T + T_2 - B^T T_3^T, \quad \Gamma_{33} = \tau Z + T_3 + T_3^T.$$

The proof of this lemma is presented in [6], using the following Lyapunov–Krasovskii functional.

$$V(x(t)) = x^T(t)Px(t) + \int_{t-d(t)}^{t} x^T(s)Qx(s) + \int_{-\tau}^{0}\int_{t+\theta}^{t} \dot{x}^T(s)Z\dot{x}(s)ds\,d\theta. \tag{19}$$

**Lemma 4** [7]: Without loss of generality, suppose that $x = 0$ is the equilibrium point of integer and fractional-order time-delay systems

$$\dot{x} = f(x(t), x(t-\tau)), \quad \tau \in [0, \infty). \tag{20}$$

$$_{t_0}^C D_t^\alpha = f(x(t), x(t-\tau)), \quad \tau \in [0, \infty), \alpha \in (0,1). \tag{21}$$

If there exists a Lyapunov–Krasovskii functional in the form

$$V_I(t) = V(x(t)) + \int_{t-\tau}^{t} g(x(s))ds \tag{22}$$

for the system (20) such that $\dot{V}_I(t)$ is negative definite and $V(x(t))$ is a convex function with respect to vector $x$, then the equilibrium point $x = 0$ of the system (21) is asymptotically stable.

**Remark 1**: For given scalars $\tau > 0$ and $\mu < 1$, the following certain fractional-order system

$$\begin{cases} D^\alpha x(t) = Ax(t) + Bx(t-d(t)), & t > 0 \\ x(t) = \phi(t), & t \in [-\tau, 0] \end{cases}. \tag{23}$$

with fixed matrices $A$ and $B$ and a time-varying state delay $d(t)$ satisfying (4) and (5) is asymptotically stable for any $0 < \alpha < 1$ if there exist $P = P^T > 0$, $Q = Q^T \geq 0$, $Z = Z^T > 0$ and appropriately dimensioned matrices $N_i$ and $T_i$ ($i = 1,2,3$) such that the LMI constraint (17) holds.

**Proof.** As Lyapunov–Krasovskii functional (19) is in the form of (22) of Lemma 4, LMI constraint (17) of Lemma 3 can also stabilize the fractional-order system (23) ∎.

## 3. Main results

In this section first, a new robust stability condition is derived for interval delay system (1) using which an LMI approach is proposed for designing a dynamic output feedback control law to robustly stabilize it.

### 3.1. *Robust stability*

In this subsection, a robust stability sufficient condition is established for the asymptotic stability of the system (1) with $u(t) \equiv 0$.

**Theorem 1**: For given scalars $\tau > 0$ and $\mu < 1$, fractional-order interval system (1), with $0 < \alpha < 1$, $A \in A_I$, $B \in B_I$, together with and a time-varying state delay $d(t)$ satisfying (4) and (5) is asymptotically stable if there exist $P = P^T > 0$, $Q = Q^T \geq 0$, $Z = Z^T > 0$ and appropriately dimensioned matrices $N_i$ and $T_i$ ($i = 1,2,3$) such that the following LMI holds:

$$\begin{bmatrix} \phi & M^T \\ M & -\eta I \end{bmatrix} < 0, \tag{24}$$

in which

$$\phi = \begin{bmatrix} Q + N_1 + N_1^T & N_2^T - N_1 & P + N_3^T + T_1 & \tau N_1 \\ \bullet & -(1-\mu)Q - N_2 - N_2^T & -N_3^T + T_2 & \tau N_2 \\ \bullet & \bullet & \tau Z + T_3 + T_3^T & \tau N_3 \\ \bullet & \bullet & \bullet & -\tau Z \end{bmatrix}$$
$$+ sym \left\{ \begin{bmatrix} -T_1 A_0 & -T_1 B_0 & \mathbf{0} & \mathbf{0} \\ -T_2 A_0 & -T_2 B_0 & \mathbf{0} & \mathbf{0} \\ -T_3 A_0 & -T_3 B_0 & \mathbf{0} & \mathbf{0} \\ \mathbf{0} & \mathbf{0} & \mathbf{0} & \mathbf{0} \end{bmatrix} \right\} \tag{25}$$

$$+\eta \begin{bmatrix} R_A & \mathbf{0} & \mathbf{0} & \mathbf{0} \\ \mathbf{0} & R_B & \mathbf{0} & \mathbf{0} \\ \mathbf{0} & \mathbf{0} & \mathbf{0} & \mathbf{0} \\ \mathbf{0} & \mathbf{0} & \mathbf{0} & \mathbf{0} \end{bmatrix} \begin{bmatrix} R_A & \mathbf{0} & \mathbf{0} & \mathbf{0} \\ \mathbf{0} & R_B & \mathbf{0} & \mathbf{0} \\ \mathbf{0} & \mathbf{0} & \mathbf{0} & \mathbf{0} \\ \mathbf{0} & \mathbf{0} & \mathbf{0} & \mathbf{0} \end{bmatrix}^T, M = \begin{bmatrix} -T_1 M_A & -T_1 M_B & \mathbf{0} & \mathbf{0} \\ -T_2 M_A & -T_2 M_B & \mathbf{0} & \mathbf{0} \\ -T_3 M_A & -T_3 M_B & \mathbf{0} & \mathbf{0} \\ \mathbf{0} & \mathbf{0} & \mathbf{0} & \mathbf{0} \end{bmatrix}$$

**Proof**: According to Remark 1 system (1) is asymptotically stable if

$$\Gamma = \begin{bmatrix} \Gamma_{11} & \Gamma_{12} & \Gamma_{13} & \tau N_1 \\ \Gamma_{12}^T & \Gamma_{22} & \Gamma_{23} & \tau N_2 \\ \Gamma_{13}^T & \Gamma_{23}^T & \Gamma_{33} & \tau N_3 \\ \tau N_1^T & \tau N_2^T & \tau N_3^T & -\tau Z \end{bmatrix} < 0, \tag{26}$$

in which

$$\Gamma_{11} = Q + N_1 + N_1^T - (A_0 + M_A F_A R_A)^T T_1^T - T_1(A_0 + M_A F_A R_A),$$

$$\Gamma_{12} = N_2^T - N_1 - (A_0 + M_A F_A R_A)^T T_2^T - T_1(B_0 + M_B F_B R_B), \Gamma_{13} = P + N_3^T + T_1 - (A_0 + M_A F_A R_A)^T T_3^T, \tag{27}$$

$$\Gamma_{22} = -(1-\mu)Q - N_2 - N_2^T - T_2(B_0 + M_B F_B R_B) - (B_0 + M_B F_B R_B)^T T_2^T,$$

$$\Gamma_{23} = -N_3^T + T_2 - (B_0 + M_B F_B R_B)^T T_3^T, \Gamma_{33} = \tau Z + T_3 + T_3^T.$$

The inequality (26) can be rewritten as follows

$$\Gamma = \begin{bmatrix} \Gamma_{11} & \Gamma_{12} & \Gamma_{13} & \tau N_1 \\ \Gamma_{12}^T & \Gamma_{22} & \Gamma_{23} & \tau N_2 \\ \Gamma_{13}^T & \Gamma_{23}^T & \Gamma_{33} & \tau N_3 \\ \tau N_1^T & \tau N_2^T & \tau N_3^T & -\tau Z \end{bmatrix} = \begin{bmatrix} Q + N_1 + N_1^T & N_2^T - N_1 & P + N_3^T + T_1 & \tau N_1 \\ \bullet & -(1-\mu)Q - N_2 - N_2^T & -N_3^T + T_2 & \tau N_2 \\ \bullet & \bullet & \tau Z + T_3 + T_3^T & \tau N_3 \\ \bullet & \bullet & \bullet & -\tau Z \end{bmatrix}$$
$$+ sym \left\{ \begin{bmatrix} -T_1 A_0 & -T_1 B_0 & \mathbf{0} & \mathbf{0} \\ -T_2 A_0 & -T_2 B_0 & \mathbf{0} & \mathbf{0} \\ -T_3 A_0 & -T_3 B_0 & \mathbf{0} & \mathbf{0} \\ \mathbf{0} & \mathbf{0} & \mathbf{0} & \mathbf{0} \end{bmatrix} \right\} \tag{28}$$
$$+ sym \left\{ \begin{bmatrix} -T_1 M_A & -T_1 M_B & \mathbf{0} & \mathbf{0} \\ -T_2 M_A & -T_2 M_B & \mathbf{0} & \mathbf{0} \\ -T_3 M_A & -T_3 M_B & \mathbf{0} & \mathbf{0} \\ \mathbf{0} & \mathbf{0} & \mathbf{0} & \mathbf{0} \end{bmatrix} \begin{bmatrix} F_A & \mathbf{0} & \mathbf{0} & \mathbf{0} \\ \mathbf{0} & F_B & \mathbf{0} & \mathbf{0} \\ \mathbf{0} & \mathbf{0} & \mathbf{0} & \mathbf{0} \\ \mathbf{0} & \mathbf{0} & \mathbf{0} & \mathbf{0} \end{bmatrix} \begin{bmatrix} R_A & \mathbf{0} & \mathbf{0} & \mathbf{0} \\ \mathbf{0} & R_B & \mathbf{0} & \mathbf{0} \\ \mathbf{0} & \mathbf{0} & \mathbf{0} & \mathbf{0} \\ \mathbf{0} & \mathbf{0} & \mathbf{0} & \mathbf{0} \end{bmatrix} \right\} < 0$$

applying Lemma 2 to the third part of the right-hand side of inequality (28), the following inequality can be obtained for a scalar $\eta > 0$

$$\Gamma = \begin{bmatrix} \Gamma_{11} & \Gamma_{12} & \Gamma_{13} & \tau N_1 \\ \Gamma_{12}^T & \Gamma_{22} & \Gamma_{23} & \tau N_2 \\ \Gamma_{13}^T & \Gamma_{23}^T & \Gamma_{33} & \tau N_3 \\ \tau N_1^T & \tau N_2^T & \tau N_3^T & -\tau Z \end{bmatrix} = \begin{bmatrix} Q + N_1 + N_1^T & N_2^T - N_1 & P + N_3^T + T_1 & \tau N_1 \\ \bullet & -(1-\mu)Q - N_2 - N_2^T & -N_3^T + T_2 & \tau N_2 \\ \bullet & \bullet & \tau Z + T_3 + T_3^T & \tau N_3 \\ \bullet & \bullet & \bullet & -\tau Z \end{bmatrix}$$

$$+ sym\left\{\begin{bmatrix} -T_1 A_0 & -T_1 B_0 & 0 & 0 \\ -T_2 A_0 & -T_2 B_0 & 0 & 0 \\ -T_3 A_0 & -T_3 B_0 & 0 & 0 \\ 0 & 0 & 0 & 0 \end{bmatrix}\right\} + \eta^{-1}\begin{bmatrix} -T_1 M_A & -T_1 M_B & 0 & 0 \\ -T_2 M_A & -T_2 M_B & 0 & 0 \\ -T_3 M_A & -T_3 M_B & 0 & 0 \\ 0 & 0 & 0 & 0 \end{bmatrix}\begin{bmatrix} -T_1 M_A & -T_1 M_B & 0 & 0 \\ -T_2 M_A & -T_2 M_B & 0 & 0 \\ -T_3 M_A & -T_3 M_B & 0 & 0 \\ 0 & 0 & 0 & 0 \end{bmatrix}^T \quad (29)$$

$$+ \eta \begin{bmatrix} R_A & 0 & 0 & 0 \\ 0 & R_B & 0 & 0 \\ 0 & 0 & 0 & 0 \\ 0 & 0 & 0 & 0 \end{bmatrix}\begin{bmatrix} R_A & 0 & 0 & 0 \\ 0 & R_B & 0 & 0 \\ 0 & 0 & 0 & 0 \\ 0 & 0 & 0 & 0 \end{bmatrix}^T < 0.$$

Inequality (29) is nonlinear because of several multiplications of variables. Therefore, by applying Schur complement on the second part of the right side of the latter inequality one has

$$\begin{bmatrix} \phi & M^T \\ M & -\eta I \end{bmatrix} < 0,$$

$$\phi = \begin{bmatrix} Q + N_1 + N_1^T & N_2^T - N_1 & P + N_3^T + T_1 & \tau N_1 \\ \bullet & -(1-\mu)Q - N_2 - N_2^T & -N_3^T + T_2 & \tau N_2 \\ \bullet & \bullet & \tau Z + T_3 + T_3^T & \tau N_3 \\ \bullet & \bullet & \bullet & -\tau Z \end{bmatrix}$$

$$+ sym\left\{\begin{bmatrix} -T_1 A_0 & -T_1 B_0 & 0 & 0 \\ -T_2 A_0 & -T_2 B_0 & 0 & 0 \\ -T_3 A_0 & -T_3 B_0 & 0 & 0 \\ 0 & 0 & 0 & 0 \end{bmatrix}\right\} \quad (30)$$

$$+ \eta \begin{bmatrix} R_A & 0 & 0 & 0 \\ 0 & R_B & 0 & 0 \\ 0 & 0 & 0 & 0 \\ 0 & 0 & 0 & 0 \end{bmatrix}\begin{bmatrix} R_A & 0 & 0 & 0 \\ 0 & R_B & 0 & 0 \\ 0 & 0 & 0 & 0 \\ 0 & 0 & 0 & 0 \end{bmatrix}^T, M = \begin{bmatrix} -T_1 M_A & -T_1 M_B & 0 & 0 \\ -T_2 M_A & -T_2 M_B & 0 & 0 \\ -T_3 M_A & -T_3 M_B & 0 & 0 \\ 0 & 0 & 0 & 0 \end{bmatrix},$$

which is equivalent to LMI in (24), and it completes the proof. ∎

### 3.2. *Robust stabilization*

The main purpose of the authors in this subsection is to design a robust dynamic output feedback controller that asymptotically stabilizes the interval FO-LTI system (1) in terms of LMIs. Hence, the following dynamic output feedback controller is presented

$$D^\alpha x_C(t) = A_C x_C(t) + B_C y(t), \quad 0 < \alpha < 1 \quad (31)$$
$$u(t) = C_C x_C(t) + D_C y(t),$$

with $x_C \in \mathcal{R}^{n_c}$, in which $n_c$ is the arbitrary order of the controller and $A_C, B_C, C_C,$ and $D_C$ are corresponding matrices to be designed.

The resulted closed-loop augmented FO-LTI system using (1) and (31) is as follows

$$D^\alpha x_{Cl}(t) = A_{Cl} x_{Cl}(t) + A_{dCl} x_{Cl}(t - d(t)), \quad 0 < \alpha < 1 \quad (32)$$

with

$$x_{Cl}(t) = \begin{bmatrix} x(t) \\ x_C(t) \end{bmatrix}, A_{Cl} = \begin{bmatrix} A & 0 \\ B_C C & A_C \end{bmatrix}, A_{dCl} = \begin{bmatrix} BD_C C & BC_C \\ 0 & 0 \end{bmatrix}. \quad (33)$$

Next, a robust stabilization theorem is established.

**Theorem 2**: For given scalars $\tau > 0$ and $\mu < 1$, closed-loop system (32), with $0 < \alpha < 1$, $A \in A_I$, $B \in B_I$, and output matrix $C$ together with and a time-varying state delay $d(t)$ satisfying (4) and (5), if there exist $P = P^T > 0$, $Q = Q^T \geq 0$, $Z = Z^T > 0$ and appropriately dimensioned matrices $N_i$, ($i = 1,2,3$) and $W_j$, ($j = 1,2,3,4$) and matrix $T = T^T$ in the form of

$$T = diag(T_S, T_C), T_S \in R^{n \times n}, T_C \in R^{n_c \times n_c}, \tag{34}$$

such that the following LMI constrain become feasible

$$\phi + \eta \Sigma_1^T \Sigma_1 + \eta^{-1} \Sigma_2^T \Sigma_2 < 0, \tag{35}$$

in which

$$\phi = sym \left\{ \begin{bmatrix} N_1 + Q & -N_1 & P + T & \tau N_1 \\ N_2 & -N_2 - (1-\mu)Q & T & \tau N_2 \\ N_3 & -N_3 & T + \tau Z & \tau N_3 \\ 0 & 0 & 0 & -\tau Z \end{bmatrix} \right\} +$$

$$sym \left\{ \begin{bmatrix} -A_0 T_S & 0 & -B_0 W_4 & -B_0 W_3 & 0 & 0 & 0 & 0 \\ -W_2 & -W_1 & 0 & 0 & 0 & 0 & 0 & 0 \\ -A_0 T_S & 0 & -B_0 W_4 & -B_0 W_3 & 0 & 0 & 0 & 0 \\ -W_2 & -W_1 & 0 & 0 & 0 & 0 & 0 & 0 \\ -A_0 T_S & 0 & -B_0 W_4 & -B_0 W_3 & 0 & 0 & 0 & 0 \\ -W_2 & -W_1 & 0 & 0 & 0 & 0 & 0 & 0 \\ 0 & 0 & 0 & 0 & 0 & 0 & 0 & 0 \\ 0 & 0 & 0 & 0 & 0 & 0 & 0 & 0 \end{bmatrix} \right\},$$

$$\Sigma_1 = \begin{bmatrix} -M_A & 0 & -M_B & -M_B & 0 & 0 & 0 & 0 \\ 0 & 0 & 0 & 0 & 0 & 0 & 0 & 0 \\ -M_A & 0 & -M_B & -M_B & 0 & 0 & 0 & 0 \\ 0 & 0 & 0 & 0 & 0 & 0 & 0 & 0 \\ -M_A & 0 & -M_B & -M_B & 0 & 0 & 0 & 0 \\ 0 & 0 & 0 & 0 & 0 & 0 & 0 & 0 \\ 0 & 0 & 0 & 0 & 0 & 0 & 0 & 0 \\ 0 & 0 & 0 & 0 & 0 & 0 & 0 & 0 \end{bmatrix}^T, \tag{36}$$

$$\Sigma_2 = \begin{bmatrix} R_A T_S & 0 & 0 & 0 & 0 & 0 & 0 & 0 \\ 0 & 0 & 0 & 0 & 0 & 0 & 0 & 0 \\ 0 & 0 & R_B D_C C T_S & 0 & 0 & 0 & 0 & 0 \\ 0 & 0 & 0 & R_B C_C T_C & 0 & 0 & 0 & 0 \\ 0 & 0 & 0 & 0 & 0 & 0 & 0 & 0 \\ 0 & 0 & 0 & 0 & 0 & 0 & 0 & 0 \\ 0 & 0 & 0 & 0 & 0 & 0 & 0 & 0 \\ 0 & 0 & 0 & 0 & 0 & 0 & 0 & 0 \end{bmatrix},$$

then, the dynamic output feedback controller parameters of

$$A_C = T_C^{-1} W_1, B_C = T_C^{-1} W_2, C_C = B^{\uparrow} T_S^{-1} W_3, D_C = B^{\uparrow} T_S^{-1} W_4, \tag{37}$$

make the closed-loop system (32) asymptotically stable.

**Proof**: The closed-loop system (32) can be considered as follows

$$D^\alpha x_{Cl}(t) = A_{Cl} x_{Cl}(t) + A_{dCl} x_{Cl}(t - d(t)), \quad 0 < \alpha < 1, \tag{38}$$

with

$$x_{Cl}(t) = \begin{bmatrix} x(t) \\ x_C(t) \end{bmatrix}, A_{Cl} = A_{0Cl} + A_{\Delta Cl}, A_{dCl} = A_{0dCl} + A_{\Delta dCl}, A_{0Cl} = \begin{bmatrix} A_0 & 0 \\ B_C C & A_C \end{bmatrix},$$

$$A_{\Delta Cl} = \begin{bmatrix} M_A F_A R_A & 0 \\ 0 & 0 \end{bmatrix}, A_{0dCl} = \begin{bmatrix} B D_C C & B C_C \\ 0 & 0 \end{bmatrix}, A_{\Delta dCl} = \begin{bmatrix} M_B F_B R_B D_C C & M_B F_B R_B C_C \\ 0 & 0 \end{bmatrix}. \tag{39}$$

According to Remark 1 the closed-loop uncertain system (32) is asymptotically stable if

$$\Gamma = \begin{bmatrix} \Gamma'_{11} & \Gamma'_{12} & \Gamma'_{13} & \tau N'_1 \\ \Gamma'^T_{12} & \Gamma'_{22} & \Gamma'_{23} & \tau N'_2 \\ \Gamma'^T_{13} & \Gamma'^T_{23} & \Gamma'_{33} & \tau N'_3 \\ \tau N'^T_1 & \tau N'^T_2 & \tau N'^T_3 & -\tau Z' \end{bmatrix} < 0, \tag{40}$$

in which we have

$\Gamma'_{11} = Q' + N'_1 + N'^T_1 - sym\{T_1(A_{0Cl} + A_{\Delta Cl})\}$,

$\Gamma_{12} = N_2^T - N_1 - (A_{0Cl} + A_{\Delta Cl})^T T_2^T - T_1(A_{0dCl} + A_{\Delta dCl})$, (41)

$\Gamma_{13} = P + N_3^T + T_1 - (A_{0Cl} + A_{\Delta Cl})^T T_3^T$, $\Gamma_{22} = -(1-\mu)Q - N_2 - N_2^T - sym\{T_2(A_{0dCl} + A_{\Delta dCl})\}$,

$\Gamma_{23} = -N_3^T + T_2 - (A_{0dCl} + A_{\Delta dCl})^T T_3^T$, $\Gamma_{33} = \tau Z + T_3 + T_3^T$,

and $Q' = Q'^T \geq 0$, $Z' = Z'^T > 0$, $P' = P'^T > 0$, $T'_i$, and $N'_i$ ($i = 1,2,3$) are matrices with appropriate dimensions. By assuming $T'_i = T', i = 1,2,3$ in the form of (34), i.e. $T' = T'^T = diag(T_S, T_C), T_S \in R^n, T_C \in R^{nc}$, and pre- and post-multiplying inequality (40) by $diag(T'^{-1}, T'^{-1}, T'^{-1}, T'^{-1})$ one has

$$\begin{bmatrix} T'^{-1} & 0 & 0 & 0 \\ 0 & T'^{-1} & 0 & 0 \\ 0 & 0 & T'^{-1} & 0 \\ 0 & 0 & 0 & T'^{-1} \end{bmatrix} \times \begin{bmatrix} \Gamma'_{11} & \Gamma'_{12} & \Gamma'_{13} & \tau N'_1 \\ \Gamma'^T_{12} & \Gamma'_{22} & \Gamma'_{23} & \tau N'_2 \\ \Gamma'^T_{13} & \Gamma'^T_{23} & \Gamma'_{33} & \tau N'_3 \\ \tau N'^T_1 & \tau N'^T_2 & \tau N'^T_3 & -\tau Z' \end{bmatrix} \times \begin{bmatrix} T'^{-1} & 0 & 0 & 0 \\ 0 & T'^{-1} & 0 & 0 \\ 0 & 0 & T'^{-1} & 0 \\ 0 & 0 & 0 & T'^{-1} \end{bmatrix} < 0$$

$$\Rightarrow \begin{bmatrix} \Gamma_{11} & \Gamma_{12} & \Gamma_{13} & \tau N_1 \\ \Gamma^T_{12} & \Gamma_{22} & \Gamma_{23} & \tau N_2 \\ \Gamma^T_{13} & \Gamma^T_{23} & \Gamma_{33} & \tau N_3 \\ \tau N^T_1 & \tau N^T_2 & \tau N^T_3 & -\tau Z \end{bmatrix} < 0 \tag{42}$$

$\Gamma'_{11} = T'^{-1}Q'T'^{-1} + T'^{-1}N'_1 T'^{-1} + T'^{-1} N'^T_1 T'^{-1} - sym\{(A_{0Cl} + A_{\Delta Cl})T'^{-1}\}$

$\Gamma'_{12} = T'^{-1}N'^T_2 T'^{-1} - T'^{-1}N'_1 T'^{-1} - T'^{-1}(A_{0Cl} + A_{\Delta Cl})^T - (A_{0dCl} + A_{\Delta dCl})T'^{-1}$

$\Gamma'_{13} = T'^{-1}P'T'^{-1} + T'^{-1}N'^T_3 T'^{-1} + T'^{-1} - T'^{-1}(A_{0Cl} + A_{\Delta Cl})^T$

$\Gamma'_{22} = -(1-\mu)T'^{-1}Q'T'^{-1} - T'^{-1}N'_2 T'^{-1} - T'^{-1}N'^T_2 T'^{-1} - sym\{(A_{0dCl} + A_{\Delta dCl})T'^{-1}\}$

$\Gamma'_{23} = -T'^{-1}N'^T_3 T'^{-1} + T'^{-1} - T'^{-1}(A_{0dCl} + A_{\Delta dCl})^T$, $\Gamma'_{33} = \tau T'^{-1} Z'^{T'^{-1}} + 2T'^{-1}$,

According to the symmetry of the matrix $T'$, the following matrices can be defined

$T = T'^{-1} = T^T$, $\quad Q = T'^{-1}Q'T'^{-1} = Q^T \geq 0$, $\quad N_1 = T'^{-1}N'_1 T'^{-1}$, $\quad N_2 = T'^{-1}N'_2 T'^{-1}$, (43)

$N_3 = T'^{-1}N'_3 T'^{-1}$, $\quad P = T'^{-1}P'^{T'^{-1}} = P^T > 0$, $\quad Z = T'^{-1}Z'^{T'^{-1}} = Z^T > 0$.

Therefore, inequality (42) can be rewritten as

$$\begin{bmatrix} \Gamma_{11} & \Gamma_{12} & \Gamma_{13} & \tau N_1 \\ \Gamma_{12}^T & \Gamma_{22} & \Gamma_{23} & \tau N_2 \\ \Gamma_{13}^T & \Gamma_{23}^T & \Gamma_{33} & \tau N_3 \\ \tau N_1^T & \tau N_2^T & \tau N_3^T & -\tau Z \end{bmatrix} = sym\left\{\begin{bmatrix} N_1+Q & -N_1 & P+T & \tau N_1 \\ N_2 & -N_2-(1-\mu)Q & T & \tau N_2 \\ N_3 & -N_3 & T+\tau Z & \tau N_3 \\ 0 & 0 & 0 & -\tau Z \end{bmatrix}\right\}$$

$$+ sym\left\{\begin{bmatrix} -A_0 T_S & 0 & -B_0 D_C C T_S & -B_0 C_C T_C & 0 & 0 & 0 & 0 \\ -B_C C T_S & -A_C T_C & 0 & 0 & 0 & 0 & 0 & 0 \\ -A_0 T_S & 0 & -B_0 D_C C T_S & -B_0 C_C T_C & 0 & 0 & 0 & 0 \\ -B_C C T_S & -A_C T_C & 0 & 0 & 0 & 0 & 0 & 0 \\ -A_0 T_S & 0 & -B_0 D_C C T_S & -B_0 C_C T_C & 0 & 0 & 0 & 0 \\ -B_C C T_S & -A_C T_C & 0 & 0 & 0 & 0 & 0 & 0 \\ 0 & 0 & 0 & 0 & 0 & 0 & 0 & 0 \\ 0 & 0 & 0 & 0 & 0 & 0 & 0 & 0 \end{bmatrix}\right\}$$

$$+ sym\left\{\begin{bmatrix} -M_A & 0 & -M_B & -M_B & 0 & 0 & 0 & 0 \\ 0 & 0 & 0 & 0 & 0 & 0 & 0 & 0 \\ -M_A & 0 & -M_B & -M_B & 0 & 0 & 0 & 0 \\ 0 & 0 & 0 & 0 & 0 & 0 & 0 & 0 \\ -M_A & 0 & -M_B & -M_B & 0 & 0 & 0 & 0 \\ 0 & 0 & 0 & 0 & 0 & 0 & 0 & 0 \\ 0 & 0 & 0 & 0 & 0 & 0 & 0 & 0 \\ 0 & 0 & 0 & 0 & 0 & 0 & 0 & 0 \end{bmatrix} \times \begin{bmatrix} F_A & 0 & 0 & 0 & 0 & 0 & 0 & 0 \\ 0 & 0 & 0 & 0 & 0 & 0 & 0 & 0 \\ 0 & 0 & F_B & 0 & 0 & 0 & 0 & 0 \\ 0 & 0 & 0 & F_B & 0 & 0 & 0 & 0 \\ 0 & 0 & 0 & 0 & 0 & 0 & 0 & 0 \\ 0 & 0 & 0 & 0 & 0 & 0 & 0 & 0 \\ 0 & 0 & 0 & 0 & 0 & 0 & 0 & 0 \\ 0 & 0 & 0 & 0 & 0 & 0 & 0 & 0 \end{bmatrix}\right.$$

(44)

$$\left.\times \begin{bmatrix} R_A T_S & 0 & 0 & 0 & 0 & 0 & 0 & 0 \\ 0 & 0 & 0 & 0 & 0 & 0 & 0 & 0 \\ 0 & 0 & R_B D_C C T_S & 0 & 0 & 0 & 0 & 0 \\ 0 & 0 & 0 & R_B C_C T_C & 0 & 0 & 0 & 0 \\ 0 & 0 & 0 & 0 & 0 & 0 & 0 & 0 \\ 0 & 0 & 0 & 0 & 0 & 0 & 0 & 0 \\ 0 & 0 & 0 & 0 & 0 & 0 & 0 & 0 \\ 0 & 0 & 0 & 0 & 0 & 0 & 0 & 0 \end{bmatrix}\right\} < 0$$

applying Lemma 2 to the third part of the right-hand side of the latter inequality, the following inequality can be obtained for a scalar $\eta > 0$

$$sym\left\{\begin{bmatrix} -M_A & 0 & -M_B & -M_B & 0 & 0 & 0 & 0 \\ 0 & 0 & 0 & 0 & 0 & 0 & 0 & 0 \\ -M_A & 0 & -M_B & -M_B & 0 & 0 & 0 & 0 \\ 0 & 0 & 0 & 0 & 0 & 0 & 0 & 0 \\ -M_A & 0 & -M_B & -M_B & 0 & 0 & 0 & 0 \\ 0 & 0 & 0 & 0 & 0 & 0 & 0 & 0 \\ 0 & 0 & 0 & 0 & 0 & 0 & 0 & 0 \\ 0 & 0 & 0 & 0 & 0 & 0 & 0 & 0 \end{bmatrix} \times \begin{bmatrix} F_A & 0 & 0 & 0 & 0 & 0 & 0 & 0 \\ 0 & 0 & 0 & 0 & 0 & 0 & 0 & 0 \\ 0 & 0 & F_B & 0 & 0 & 0 & 0 & 0 \\ 0 & 0 & 0 & F_B & 0 & 0 & 0 & 0 \\ 0 & 0 & 0 & 0 & 0 & 0 & 0 & 0 \\ 0 & 0 & 0 & 0 & 0 & 0 & 0 & 0 \\ 0 & 0 & 0 & 0 & 0 & 0 & 0 & 0 \\ 0 & 0 & 0 & 0 & 0 & 0 & 0 & 0 \end{bmatrix}\right.$$

(45)

$$\left.\times \begin{bmatrix} R_A T_S & 0 & 0 & 0 & 0 & 0 & 0 & 0 \\ 0 & 0 & 0 & 0 & 0 & 0 & 0 & 0 \\ 0 & 0 & R_B D_C C T_S & 0 & 0 & 0 & 0 & 0 \\ 0 & 0 & 0 & R_B C_C T_C & 0 & 0 & 0 & 0 \\ 0 & 0 & 0 & 0 & 0 & 0 & 0 & 0 \\ 0 & 0 & 0 & 0 & 0 & 0 & 0 & 0 \\ 0 & 0 & 0 & 0 & 0 & 0 & 0 & 0 \\ 0 & 0 & 0 & 0 & 0 & 0 & 0 & 0 \end{bmatrix}\right\} \leq \eta \Sigma_1^T \Sigma_1 + \eta^{-1} \Sigma_2^T \Sigma_2,$$

in which we have

$$\Sigma_1 = \begin{bmatrix} -M_A & 0 & -M_B & -M_B & 0 & 0 & 0 & 0 \\ 0 & 0 & 0 & 0 & 0 & 0 & 0 & 0 \\ -M_A & 0 & -M_B & -M_B & 0 & 0 & 0 & 0 \\ 0 & 0 & 0 & 0 & 0 & 0 & 0 & 0 \\ -M_A & 0 & -M_B & -M_B & 0 & 0 & 0 & 0 \\ 0 & 0 & 0 & 0 & 0 & 0 & 0 & 0 \\ 0 & 0 & 0 & 0 & 0 & 0 & 0 & 0 \\ 0 & 0 & 0 & 0 & 0 & 0 & 0 & 0 \end{bmatrix}^T,$$

(46)

$$\Sigma_2 = \begin{bmatrix} R_A T_S & 0 & 0 & 0 & 0 & 0 & 0 & 0 \\ 0 & 0 & 0 & 0 & 0 & 0 & 0 & 0 \\ 0 & 0 & R_B D_C C T_S & 0 & 0 & 0 & 0 & 0 \\ 0 & 0 & 0 & R_B C_C T_C & 0 & 0 & 0 & 0 \\ 0 & 0 & 0 & 0 & 0 & 0 & 0 & 0 \\ 0 & 0 & 0 & 0 & 0 & 0 & 0 & 0 \\ 0 & 0 & 0 & 0 & 0 & 0 & 0 & 0 \\ 0 & 0 & 0 & 0 & 0 & 0 & 0 & 0 \end{bmatrix}.$$

Substituting (45) in (44), yields into

$$\begin{bmatrix} \Gamma_{11} & \Gamma_{12} & \Gamma_{13} & \tau N_1 \\ \Gamma_{12}^T & \Gamma_{22} & \Gamma_{23} & \tau N_2 \\ \Gamma_{13}^T & \Gamma_{23}^T & \Gamma_{33} & \tau N_3 \\ \tau N_1^T & \tau N_2^T & \tau N_3^T & -\tau Z \end{bmatrix} = sym\left\{\begin{bmatrix} N_1 + Q & -N_1 & P + T & \tau N_1 \\ N_2 & -N_2 - (1-\mu)Q & T & \tau N_2 \\ N_3 & -N_3 & T + \tau Z & \tau N_3 \\ 0 & 0 & 0 & -\tau Z \end{bmatrix}\right\}$$

$$+ sym\left\{\begin{bmatrix} -A_0 T_S & 0 & -B_0 D_C C T_S & -B_0 C_C T_C & 0 & 0 & 0 & 0 \\ -B_C C T_S & -A_C T_C & 0 & 0 & 0 & 0 & 0 & 0 \\ -A_0 T_S & 0 & -B_0 D_C C T_S & -C_C T_C & 0 & 0 & 0 & 0 \\ -B_C C T_S & -A_C T_C & 0 & 0 & 0 & 0 & 0 & 0 \\ -A_0 T_S & 0 & -B_0 D_C C T_S & -C_C T_C & 0 & 0 & 0 & 0 \\ -B_C C T_S & -A_C T_C & 0 & 0 & 0 & 0 & 0 & 0 \\ 0 & 0 & 0 & 0 & 0 & 0 & 0 & 0 \\ 0 & 0 & 0 & 0 & 0 & 0 & 0 & 0 \end{bmatrix}\right\} + \eta \Sigma_1^T \Sigma_1 + \eta^{-1} \Sigma_2^T \Sigma_2 < 0.$$

(47)

Inequality (47) is nonlinear due to various multiplications of variables. Hence, by applying Schur complement on $\eta^{-1}\Sigma_2^T\Sigma_2$, and changing variables as follows

$$W_1 = T_C A_C, \qquad W_2 = T_C B_C, \qquad W_3 = T_S B C_C, \qquad W_4 = T_S B D_C,$$

(48)

one can obtain linear matrix inequality (35) with parameters in (36). ∎

**Corollary 1**: Although Theorem 1 and Theorem 2 are respectively allocated to robust stability and stabilization of uncertain FO-LTI systems of form (1), the proposed method can be easily used for the case of certain systems by solving the LMI constraints $\phi < 0$ in these theorems, respectively.

**Proof:** The proof is straightforward by assuming $A_{\Delta Cl} = \mathbf{0}$ in proof procedure of Theorem 1 and Theorem 2.

## 4. Numerical examples

In this section, some numerical examples are given to demonstrate the applicability of the proposed method. In this paper, we use YALMIP parser [24] and SeDuMi [25] solver in Matlab tool [26] in order to assess the feasibility of the proposed constraints to obtain the controller parameters.

4.1. Example *1*

In [1] robust stability of the fractional-order interval system

$$G(s) = \frac{[1.3,1.7]s^{0.3}+[1.4,1.6]}{[1.5,2.5]s^{0.6}+[2.5,3.5]s^{0.3}+[1.5,2.5]}e^{-0.1s}, \tag{49}$$

with fractional-order $PI$ controller $C(s) = 2 + 0.5s^{-0.3}$, proposed in [15], is checked. The aim of this subsection is to check robust stability of the closed-loop delayed system using proposed LMI constraints in Theorem 1. The pseudo-state space representation of form (1) for the given system is as follows

$$\underline{A} = \begin{bmatrix} -2.3333 & 1 \\ -1.6667 & 0 \end{bmatrix}, \overline{A} = \begin{bmatrix} -1 & 1 \\ -0.6000 & 0 \end{bmatrix}, \underline{B} = \begin{bmatrix} 0.52 \\ 0.56 \end{bmatrix}, \overline{B} = \begin{bmatrix} 1.1333 \\ 1.0667 \end{bmatrix}, \alpha = 0.3. \tag{50}$$

Moreover, the pseudo-state space representation of the given controller is

$$A_C = 0, B_C = 1, C_C = 0.5, D_C = 2, \tag{51}$$

therefore, closed-loop system of form (32) can be represented by following parameters:

$$\underline{A_{cl}} = \begin{bmatrix} -2.3333 & 1 & 0 \\ -1.6667 & 0 & 0 \\ 1 & 0 & 0 \end{bmatrix}, \overline{A_{cl}} = \begin{bmatrix} -1 & 1 & 0 \\ -0.6000 & 0 & 0 \\ 1 & 0 & 0 \end{bmatrix},$$
$$\underline{A_{dcl}} = \begin{bmatrix} 1.0400 & 0 & 0.2600 \\ 1.1200 & 0 & 0.2800 \\ 0 & 0 & 0 \end{bmatrix}, \overline{A_{dcl}} = \begin{bmatrix} 2.2666 & 0 & 0.5666 \\ 2.1334 & 0 & 0.5333 \\ 0 & 0 & 0 \end{bmatrix}, d(t) = 0.1. \tag{52}$$

Using Theorem 1 following parameters can be obtained, which illustrate the robust stability of the uncertain system (49), which has been concluded in [1], by calculating a bound on the poles of fractional-order interval systems and extending the concept of the value set and zero exclusion principle to these systems.

$$\eta = 0.0014 > 0, \quad P = \begin{bmatrix} 14.0163 & -16.3965 & -1.7118 \\ -16.3965 & 20.9606 & 0.6398 \\ -1.7118 & 0.6398 & 1.2590 \end{bmatrix} > 0,$$
$$Q = \begin{bmatrix} 13.4193 & -13.0917 & -3.4991 \\ -13.0917 & 18.3259 & -0.5569 \\ -3.4991 & -0.5569 & 3.8213 \end{bmatrix} \geq 0, Z = \begin{bmatrix} 27.2212 & -24.7145 & -8.2742 \\ -24.7145 & 39.1922 & -4.9566 \\ -8.2742 & -4.9566 & 12.7109 \end{bmatrix} > 0,$$
$$T_1 = \begin{bmatrix} -17.2581 & 16.1175 & 2.7759 \\ 16.1175 & -24.5352 & -2.3622 \\ 2.7759 & -2.3622 & 0.0097 \end{bmatrix}, T_2 = \begin{bmatrix} 6.8306 & 0.6292 & 0.0485 \\ 0.6292 & 7.3876 & 4.6803 \\ 0.0485 & 4.6803 & -0.5992 \end{bmatrix}, \tag{53}$$
$$T_3 = \begin{bmatrix} -9.5420 & 5.5947 & 0.9088 \\ 5.5947 & -12.9149 & -1.2126 \\ 0.9088 & -1.2126 & -3.8069 \end{bmatrix}, N_1 = \begin{bmatrix} -6.1919 & -0.9171 & 2.4553 \\ -0.9171 & -6.2318 & 2.7583 \\ 2.4553 & 2.7583 & -8.5485 \end{bmatrix},$$
$$N_2 = \begin{bmatrix} 1.1762 & 1.4902 & 1.4902 \\ 1.4902 & 2.5406 & -0.9299 \\ -1.4340 & -0.9299 & 4.9673 \end{bmatrix}, N_3 = \begin{bmatrix} 9.6722 & -4.5325 & -2.8009 \\ -4.5325 & 2.0974 & 3.3851 \\ -2.8009 & 3.3851 & 1.6671 \end{bmatrix}.$$

4.2. Example 2 (robust stabilization)

The dynamic output feedback stabilization problem of the uncertain fractional-order system of Example 1 in the form of (1) is considered with $A \in A_I = [\underline{A}, \overline{A}]$ and $B \in B_I = [\underline{B}, \overline{B}]$ presented in (50), and the time-varying delay $d(t)$ is considered as follows

$$d(t) = 0.15(\sin(t) + 1)(1 - e^{-t}) \Rightarrow \tau = 0.25, \mu = 0.15 \tag{54}$$

According to Theorem 2, it can be concluded that this uncertain fractional-order system is asymptotically stabilizable utilizing the obtained dynamic output feedback controllers in the form of (31), with controller orders $n_c = 0,1,2$, tabulated in Table 1.

The time response of the uncertain closed-loop FO-LTI system of form (32), consisting of a random system in the interval (50) and the obtained controller with $n_c = 0$ (static controller) is illustrated in Figure 1 where all the states asymptotically converge to zero. The eigenvalues of $A_{Cl}$, for some random systems in the above interval, and stability boundaries $\pm \alpha \pi/2$ are demonstrated in Figure 2, where all of the eigenvalues of $A_{Cl}$ are located in the stability region. It is obvious from Figure 1 and Figure 2 that stabilizing of the interval FO-LTI system with time-varying delay is possible even with proposed static controller with $n_c = 0$.

Table 1. The obtained controller parameters for Example 2 using Theorem 2.

| $n_c$ | $A_C$ | $B_C$ | $C_C$ | $D_C$ |
|---|---|---|---|---|
| 0 | 0 | 0 | 0 | $-1.4215$ |
| 1 | $-24.0413$ | $-0.2088$ | $-0.0061$ | $-1.6597$ |
| 2 | $\begin{bmatrix} -21.4575 & -0.1151 \\ -0.1114 & -21.4721 \end{bmatrix}$ | $\begin{bmatrix} -0.1930 \\ -0.2219 \end{bmatrix}$ | $\begin{bmatrix} -0.0461 \\ -0.0330 \end{bmatrix}^T$ | $-1.5227$ |

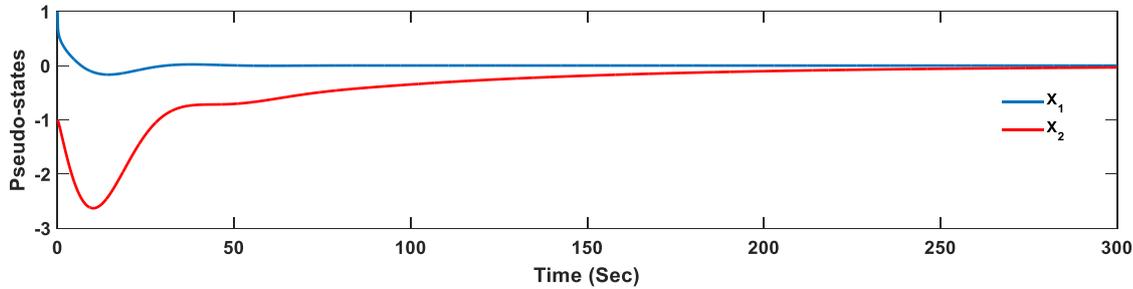

Figure 1. Pseudo-state trajectory of closed-loop FO-LTI system of form (32), via obtained controller with $n_C = 0$.

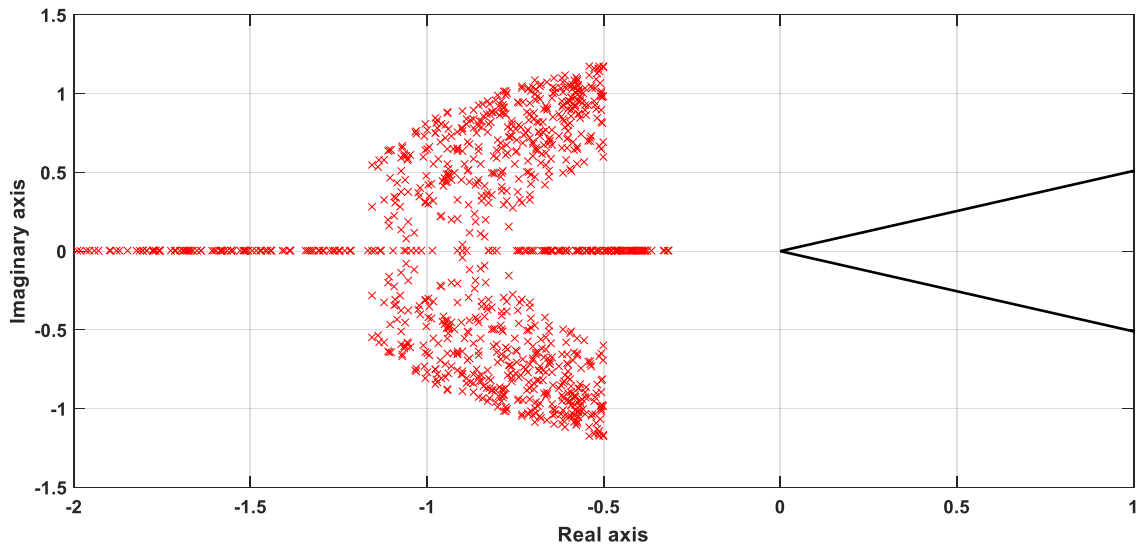

Figure 2. The location of eigenvalues of the uncertain closed-loop system via obtained output feedback controller in Example 2 with $n_c = 0$.

## 5. Conclusion

This paper has solved the problem of stability and stabilization of interval fractional-order systems with time-varying delay, where the elements of the systems pseudo-state space matrices are uncertain parameters that each adopts a value in a real interval. The time-varying delay also offers more generality compared with time-constant one which has been adopted in previous works. Utilizing various lemmas, the stability and stabilization theorems are proposed in the form of LMIs, which is more suitable to check due to various existing efficient convex optimization parsers and solvers. Eventually, two numerical examples have shown the effectiveness of proposed robust stability and stabilization theorems.